\documentclass[11pt]{article}
\usepackage{amsmath}
\usepackage{amsfonts}
\usepackage{amssymb}
\usepackage{graphicx,color}
\usepackage{subfigure}
\usepackage{hyperref}
\usepackage[usenames,dvipsnames,svgnames,table]{xcolor}
\usepackage{enumerate}
\usepackage[left=2cm,top=2.5cm,right=2cm,bottom=2cm]{geometry}

\def\be{\begin{equation}}
\def\ee{\end{equation}}
\def\ba{\begin{array}}
\def\ea{\end{array}}

\newcommand{\bea}{\begin{eqnarray}}
\newcommand{\eea}{\end{eqnarray}}

\def \be  {\begin{equation}}
\def \ee  {\end{equation}}
\def \ba  {\begin{eqnarray}}
\def \ea  {\end{eqnarray}}

\begin{document}
\title{Hybrid Natural Inflation} 
\author{Graham G. Ross$^{a}$, Gabriel Germ\'an$^{a}$\footnote{On sabbatical leave from
Instituto de Ciencias F\'isicas, Universidad Nacional Aut\'onoma de M\'exico, UNAM.
}, J. Alberto V\'azquez$^{b}$\\
\\
{\normalsize \textit{$^a$Rudolf Peierls Centre for Theoretical Physics,} }\\
{\normalsize \textit{University of Oxford, 1 Keble Road, Oxford, OX1 3NP, UK}}\\
\\
{\normalsize \textit{$^b$Brookhaven National Laboratory, }}\\
{\normalsize \textit{2 Center Road, Upton, NY 11973, USA.}}\\
\\
}
\maketitle

\begin{abstract}
We construct two simple effective field theory versions of {\it Hybrid Natural Inflation (HNI)} that illustrate the range of its phenomenological implications. The resulting inflationary sector potential, $V=\Delta^4(1+a\cos(\phi/f))$, arises naturally, with the inflaton field a pseudo-Nambu-Goldstone boson. The end of inflation is triggered by a waterfall field and the conditions for this to happen are determined. Also of interest is the fact that the slow-roll parameter $\epsilon$ (and hence the tensor $r$) is a non-monotonic function of the field with a maximum where observables take universal  values that determines the maximum possible tensor to scalar ratio $r$. In one of the models the inflationary scale can be as low as the electroweak scale. We explore in detail the associated HNI phenomenology, taking account of the constraints from Black Hole production, and perform a detailed fit to the Planck 2015 temperature and polarisation data.
\end{abstract}
\section{Introduction} \label{Intro} 
\noindent
Among the many models proposed to implement the inflationary paradigm \cite{Guth:1980zm}, \cite{Linde:1981mu}, \cite{Albrecht:1982wi}, \cite{Lyth:1998xn}, {\it Natural Inflation} (NI)  \cite{Freese:1990rb}, \cite{Adams:1992bn}, \cite{Freese:2008if}, \cite{Freese:2014nla} is particularly appealing because its origins lie in well motivated 
physics.  In this scheme the inflaton potential has the form 
\be
V_I(\phi)=\Delta^4(1+\cos(\frac{\phi}{ f})),
\label{pot}
\ee
where the inflaton, $\phi$, is a pseudo-Goldstone boson associated with a spontaneously broken global symmetry and is thus protected from large radiative corrections to its mass.
Unfortunately, the predictions of NI are now only marginally consistent with the recent measurements \cite{Planck:2013jfk}. In addition it requires the symmetry breaking scale, $f$, to be larger than the Planck scale $M = 2.44 \times 10^{18} GeV$\footnote{ In what follows the Planck scale will be taken equal to unity.} raising doubts about the stability of the potential against higher dimensional terms\footnote{Modified schemes have been constructed with additional fields and sub-Planckian  scales of symmetry breaking but where the resulting effective scale is super-Planckian \cite{Kim:2004rp}, \cite{Dimopoulos:2005ac}; see however \cite{Palti:2015xra}, \cite{Baume:2016psm}.}.
However it is possible to construct generalised ``Hybrid Natural Inflation"  models
\cite{Ross:2009hg}-\cite{Carrillo-Gonzalez:2014tia},  that maintain the symmetry protection for the inflaton mass, are perfectly consistent with all current measurements and can avoid the need for a super-Planckian symmetry breaking scale. The inflaton potential relevant to the inflationary era now has the general form 
\be
V_I(\phi)=\Delta^4(1+a\cos(\frac{\phi}{ f})),
\label{pot1}
\ee
where $0\le a<1$. The change in the structure is because inflation ends due to a new hybrid ``waterfall" field \cite{Linde:1993cn}, $\chi$, that couples to the inflaton and ends inflation when this coupling triggers $\chi$ to develop a vacuum expectation value (vev). The appearance of the new parameter, $a$, allows for more general inflationary phenomena that can readily accommodate the Planck results and even allow for a low-scale of inflation. The waterfall field is important in the era after inflation and can lead to efficient reheating of the universe.  

Our paper is organized as follows: In Section \ref{EFT}, we construct the effective field theory (EFT) of HNI that includes the waterfall field and is valid below the scale, $\Lambda$, corresponding to the scale of the ultra-violet (UV) completion of the model. This may be the scale at which the theory becomes supersymmetric or the composite scale or even the Planck scale. Although the inflaton is protected by the underlying Goldstone symmetry from large corrections to its mass proportional to $\Lambda$, the same is not true of the waterfall field  and so there is a constraint on $\Lambda$ following from the requirement that HNI should naturally avoid fine tuning. As we discuss, there are essentially  two classes of HNI depending on the underlying symmetries of the EFT. In one class it is possible significantly to lower the scale of inflation and we discuss the limits on this scale. We also discuss how the initial conditions prior to inflation may occur and the constraints on the reheat temperature after inflation. In Section~\ref{Slow} we  write the form slow-roll (SR) parameters and observables in terms of a convenient notation. In Section \ref{approx} we consider the phenomenological implications of HNI in the sub-Planckian $f$ limit that can be analysed analytically. We construct the slow roll parameters and the associated results for both scalar and tensor density perturbations and compare them to the Planck data. We show that there is an upper bound to $r$ and  that in one class of HNI models the inflation scale may be as low as the electroweak scale. In Section \ref{num}  we perform a likelihood fit of HNI to the available data that allows us to determine the range of observables consistent with HNI. In this we do not constrain $f$ to be sub-Planckian.
Section \ref{pbh} presents a discussion of constraints on HNI coming from primordial black hole abundances bounds at the end of inflation. We also check consistency of the hierarchy of SR parameters with the usual first order power spectrum formula.  Finally, we conclude in Section~\ref{conclu}  by discussing the main results obtained in the paper coming from observational and theoretical constraints on the model.
\section{The effective field theory description of Hybrid Natural Inflation}\label{EFT}
\subsection{The simplest scheme}
Natural inflation identifies the inflaton with a Pseudo-Goldstone boson, $\phi$. The field theoretic origin of the pseudo-Goldstone mode is the phase of a complex scalar 
field, $\Phi$, such that
\be
\Phi  = (\rho  + \tilde f){e^{i\frac{\phi }{\tilde f}}},
\label{Phi}
\ee
where $\tilde f$ the scale of the Goldstone symmetry breaking and $\rho$ is the radial field that acquires a mass of $O(\tilde f)$. 
To obtain an hybrid version of NI it is necessary to have at least an additional field that in the simplest implementation can be taken as a real field, $\chi$. Then the scalar potential can be written in the form
\be
V\left( {\Phi ,\chi} \right) = {V_0}\left( {\left| \Phi  \right|, \chi } \right) + {V_1}\left( {\Phi ,\chi } \right) + {V_2}\left( \Phi\right).
\label{Pot}
\ee
The first term is invariant under the global $U(1)$ symmetry, $\Phi\rightarrow e^{i\alpha}\Phi$, and has the general structure\footnote{For simplicity we assume $V_0$ is invariant under $\chi\rightarrow -\chi$ but this can be relaxed without significantly changing the model.}
\be
V_0=-m_{\phi}^2 |\Phi|^2+\lambda |\Phi |^4+m_{\chi}^2 \chi^2 +h_1 \chi^4+h_2 |\Phi|^2\chi^2+\bar\Delta^4,
\label{Pot0}
\ee
where we have allowed for a constant term, $\bar\Delta^4$, to be present that may come from other terms in the UV completion of the model.
For positive $m_\phi^2$, $\Phi$ triggers spontaneous breaking of the $U(1)$ symmetry. In this case $\Phi$ is better parameterised by Eq. (\ref{Phi})  with
\be
\tilde f=\sqrt{{\frac{m_\phi^2}{ 2\lambda}}},\;\;m_\rho^2=2m_\phi^2 ,
\label{f2}
\ee
where $\tilde f$ is the vev of $\Phi$ and $\phi$ is the massless Goldstone boson associated with this breaking. 
The remaining terms in Eq. (\ref{Pot}) explicitly break the $U(1)$ symmetry and generate a mass for the Goldstone mode. This mass is governed by the magnitude of the couplings in these breaking terms and for small couplings the mass will be small allowing for a flat inflationary potential. The potential $V_1(\Phi,\chi)$  is responsible for ending inflation because it generates a mass term for $\chi$ that depends on the $\phi$ vev. As the mass squared becomes negative it triggers a vev for $\chi$, reducing $V$ and ending the slow roll. The form of $V_1$ may be limited by discrete symmetries and we choose to implement a $Z_2$ symmetry, $\Phi\rightarrow -\Phi^{\dagger}$ that restricts $V_1$ to the form
\be
V_1={\frac{1}{ 2}}\delta (\Phi^2+\Phi^{\dagger 2})\chi^2=\delta\;(\rho+\tilde f)^2\cos({\frac{2\phi}{ \tilde f}})\chi^2.
\ee
We see that the $U(1)$ symmetry is broken by this term to a discrete $Z_2$ subgroup corresponding to $\alpha=\pi$. 
Since $\rho$ acquires an unsuppressed mass it plays no role in ending inflation and we will ignore it from now on.
Finally we should include the most general potential, $V_2(\Phi)$, that is consistent with the $Z_2$ symmetry. It is given by
\bea
V_2(\Phi)&=&{\frac{1}{ 2}}m_{\phi '}^2(\Phi^2+\Phi^{\dagger 2})+{\frac{1}{ 2}}\lambda (\Phi^4+\Phi^{\dagger  4})\nonumber\\
&\rightarrow&m_{\phi'}^2\tilde f^2\cos(\frac{2\phi}{ \tilde f})+ \lambda \tilde f^4\cos(\frac{4\phi}{ \tilde f}).
\label{v2}
\eea
Note that there is a minimum value for $m_{\phi'}$ and $\lambda$ that can be taken without imposing unnatural fine tuning. This is because such terms are generated by radiative 
corrections and we must include them if we claim to have a natural inflationary theory. In the absence of fine tuning these radiative terms require that\footnote{Due to the non-renormalisation theorem, there is an additional suppression  by the factor $\frac{\Lambda^2}{ m_\chi^2}$ of the radiative corrections to  $\lambda$ for the case of a supersymmetric UV completion.}
\bea
m_{\phi'}^2&\ge& \frac{\delta \Lambda^2}{ 16\pi^2},\nonumber\\
\lambda&\ge&\frac{\delta^2}{ 16\pi^2}\,.
\eea
Finally there are potentially large 
radiative corrections to the waterfall field, $\chi$, that limit how small we can take $m_\chi $.  In this case we 
must take $m_\chi^2>\alpha \Lambda^2$, where $\alpha$ is a radiative factor, $\alpha=O(h_1/16\pi^2)+
\cdot \cdot \cdot$ and $\Lambda$ is the cutoff scale for the radiative corrections mentioned above. 
\subsubsection{The inflationary era}
During inflation the waterfall field plays no roll. The explicit $U(1)$ breaking term is given by $V_2$. Taking the radiative corrections as indicative of the magnitude of the terms,  a light inflaton requires small $\delta$ and the dominant radiative correction will be to $m_{\phi'}$ with the first term of Eq. (\ref{v2})  giving a scalar potential of the HNI form, Eq. ({\ref{pot1}), with
\bea
f&=&\frac{\tilde f}{ 2},\nonumber\\
a&=&\frac{\delta \tilde f^2\Lambda^2}{ 16\pi^2 \Delta^4},\nonumber\\
\Delta^4&=&\bar\Delta^4-\frac{m_\phi^4}{ 4 \lambda}\,.
\eea
\subsubsection{The post-inflationary era}
The crucial point of HNI is that inflation ends when the change in the inflation vev triggers a negative value for the mass squared of the waterfall field $\chi$ so that, once it exceeds the square of the Hubble parameter, it runs to its minimum reducing the potential and thus ending the slow-roll inflation of $\phi$. 
  The condition for this to happen is 
\be
-\left( m_\chi^2+4h_2f^2+4\delta f^2\cos(\frac{\phi_e}{ f})\right)M^2\ge  \Delta^4,
\label{sle}
\ee
 thus it is necessary that
\be
\tilde m_\chi^2\equiv m_\chi^2+4h_2f^2+\frac{\Delta^4}{ M^2}<4\delta  f^2.
\label{massconstraint}
\ee
Inflation ends when $\phi=\phi_e$ where
\be
\cos(\frac{\phi_e}{ f})\approx -\frac{\tilde m_\chi^2}{ 4 \delta  f^2}.
\ee
Note that to avoid fine tuning between unrelated parameters there is a limit on how close $\phi_e/f$ can be  to $\pi$. Thus, if the coefficient of the cosine term in Eq. (\ref{sle}) is 10\% greater than the magnitude of the sum of the remaining terms, $\cos(\phi_e/f)\sim0.9$ corresponding to $\phi_e/f=0.86\pi$ and if the difference is only 1\% $\phi_e/f=0.95\pi$. This will be important when determining the number of e-folds of inflation below.

After inflation ends the waterfall field rolls to its minimum with $\cos(\frac{\phi}{ f})=-1$ and 
\be
V=\Delta^4(1-a)-{\bar m_\chi^4\over 4h_1},
\ee
where 
\be
\bar m_\chi^2=4\delta  \tilde f^2-4h_2 \tilde f^2-m_\chi^2.
\ee
As is usually done in inflationary models we must fine-tune to get zero cosmological constant after reheat so 
\be
\bar m_\chi^4=4h_1\Delta^4(1-a).
\ee

Below we will discuss the limits on the scale of inflation that result from the constraints on the parameters just discussed. However before doing this we construct another version of the coupling of the waterfall field that exhibits another extreme of this class of models.
\subsection{An alternative model}
The model just constructed used a $Z_2$ symmetry to restrict the couplings of the EFT. Here we choose an alternative $Z_2\times Z_2'$ symmetry that generates a different structure for the waterfall potential.  These models illustrate two extremes while the more general model built without imposing  $Z_2$ symmetries interpolates between the two models as its parameters are varied. 

To build this alternative model we first extend the model to incorporate a complex, rather than a real, scalar field $\chi\equiv \chi_R+i\chi_I$. In this case the first term in the  potential has the form 
\be
V_0(|\Phi|,|\chi|)=-m_{\phi}^2 |\Phi|^2+\lambda |\Phi |^4+m_{\chi}^2 |\chi|^2 +h_1 |\chi|^4+h_2 |\Phi|^2|\chi|^2+\bar\Delta^4,
\label{Pot2}
\ee
and is invariant under a $U(1)_\Phi\times U(1)_\chi$ symmetry. We assume that $m_\phi^2$ and $m_\chi^2$ are positive so $\Phi$ acquires a vev as before but $\chi$ does not. 

The coupling between the inflaton and the waterfall field proceeds through the term
\bea
V_1(\Phi,\chi)&=&{\delta\over 8}(\Phi^2-\Phi^{\dagger 2})(\chi^2-\chi^{\dagger 2})\nonumber\\
&=&-\delta \;\tilde f^2\;\sin({2\phi\over \tilde f})\chi_R\chi_I,
\label{wf}
\eea
which is the only such term allowed by a $Z_2\times Z_2'$ symmetry defined by $\Phi\rightarrow \Phi^{\dagger}$, $\chi\rightarrow \chi^{\dagger}$ and $\Phi\rightarrow i\Phi^{\dagger}$, $\chi\rightarrow -\chi$. Clearly this term breaks the $U(1)_\Phi$ symmetry. 

Finally there are further terms allowed by this symmetry given by
\bea 
V_2&=&{\lambda\over 2}(\Phi^4+\Phi^{\dagger 4})+{{m_{\chi'}^2}\over 2}(\chi^2+\chi^{\dagger 2})+{\lambda_\chi\over 2}(\chi^4+\chi^{\dagger 4})\nonumber\\
&=&\lambda \tilde f^4 \cos({4\phi\over \tilde f})+m_{\chi'}^2(\chi_R^2-\chi_I^2)+\lambda_\chi(\chi_R^4-6\chi_R^2\chi_I^2+\chi_I^4).
\eea
\subsubsection{The inflationary era}
One sees that there is again an inflation potential of the HNI form with 
\bea
f&=&{\tilde f\over 4},\nonumber\\
a&=&\lambda {\tilde f^4\over \Delta^4},\nonumber\\
\Delta^4&=&\bar\Delta^4-{m_\phi^4\over 4 \lambda}.
\eea
As before there are radiative corrections that limit how small the couplings can be. However the symmetries of the theory mean that there is no correction to $m_\chi'$ and so this term can be arbitrarily small. The coupling $\lambda$ does get a correction so that it has a natural lower bound given by
\be
\lambda\ge {\delta^2 \over 16\pi^2},
\ee
and, as above,  there is an additional suppression factor ${\Lambda^2\over m_\chi^2}$ for the case of a supersymmetric UV completion.
\subsubsection{The post-inflationary era}
For the case that $m_\chi'^2$ is positive the condition that the waterfall field ends inflation is  given by 
\be
\tilde m_\chi^2\equiv m_\chi^2+16h_2f^2+{\Delta^4\over M^2}<8\; \delta f^2,
\ee
because the  waterfall vevs can develop along the direction $<\chi_R>=<\chi_I>.$

Note  that the  inflaton dependence of $V_2$ is different from that in $V_1$ whereas in the first model the two terms have the same inflaton dependence. As a result the end of inflation occurs when $\phi=\phi_e$ where 
 \be
 \sin({\phi_e\over 2f})\approx{\tilde m_\chi^2\over 8\;\delta f^2}.
 \ee
 
As in the previous model there is a similar fine tuning constraint on how close $(\phi_e/f)$ can be to $\pi$ as the numerator and denominator are unrelated parameters. Note however that there is no  fine tuning restriction on  how small $\phi_e/f$ can be because it is possible the denominator is arbitrarily larger than the numerator provided $\tilde m_\chi$ is protected from acquiring a large radiative mass by a symmetry (the low $\Lambda$ case).  As we will discuss this leads to significant phenomenological implications, these two models representing two extremes in the waterfall field behaviour.
 
 In this model the condition the cosmological constant vanishes after inflation is given by
 \be
\bar m_\chi^4=16h_1\Delta^4(1-a),
\ee
where 
\be
\bar m_\chi^2=16\delta f^2-32h_2 f^2-2m_\chi^2.
\ee

\subsection{Initial conditions for inflation}
For a slow-roll inflationary period to occur the common belief is that there must initially be a horizon-size volume 
of space with a very uniform vev for the inflaton field. The problem is much more severe in the case of low-scale inflation because of the growth of the horizon size so that the constraint on homogeneity extends over a huge number of Planck scale horizon volumes. There have been several suggestions to address this question, all of them requiring some earlier period of, possibly eternal, inflation.

One possible explanation for this is that there was a previous inflationary era at, or near to, the Planck scale 
so that one needs homogeneity over only a few Planck scale horizon volumes but that these would be blown up by the initial inflationary era to be larger than the low-scale Planck volume and generate the homogeneous initial conditions necessary for low-scale inflation to  occur   \cite{Hawking:1998bn} - \cite{Turok:2000bt}. 
Of course there remains the question why the initial vev of the inflaton should be in the domain that allows for a subsequent slow-roll inflationary period. For the first waterfall field model ${\phi_0\over f}=O(1)$ and so there is no need for fine tuning of the initial vev. However for the second waterfall field model with a low scale of inflation Eq. (\ref{lowscale}) requires the initial value of ${\phi_0\over f}$ is very small. It is possible that thermal effects could drive $\phi_H$ towards the origin but this in turn requires that the effective temperature during the first stage of inflation should be less than $m_\phi\sim f$ so that $\phi$  develops a vev in this era. One can also argue that there is no need for an explanation of the initial value of ${\phi_0\over f}$ because, with random initial values, the ones leading to inflation will dominate the late-time universe. This of course leads to the need to discuss the measure determining relative probabilities but this takes us far beyond the scope of this paper. 

Another possible explanation for the initial conditions again relies on an earlier period of inflation but this time due to the universe being trapped in a false vaccuum state \cite{Vilenkin:1982de} - \cite{Zeldovich:1984vk}. Tunnelling from this state can lead to a homogeneous bubble with the appropriate initial conditions for HNI to occur. 

Yet another possibility is topological inflation \cite{Vilenkin:1994pv} in which an  horizon volume fits in a topologically stable structure (domain wall) with non-vanishing vacuum energy. As this volume inflates its extremities are no longer stable and may have the appropriate initial conditions for HNI inflation to occur \footnote { For a recent review on initial conditions for inflation obtained from scalar fields minimally coupled to General Relativity see \cite{Brandenberger:2016uzh}.}. 

However, recently the requirement of an horizon-size volume of space with a uniform vev has been questioned. Numerical studies of a scalar field coupled
to Einstein equations in 3+1 dimensions suggest \cite{East:2015ggf} that under certain circumstances an
inflationary period can result even from an initial
inhomogeneous universe dominated by gradient and kinetic
energy instead of the usual potential energy dominating
term. A possible understanding of this phenomenon could be
that the gradient and kinetic energy dilute due to the expansion until the
vacuum energy dominates starting inflation as usually
understood.
\subsection{Reheating}
\subsubsection{Model 1} 
In this model inflation ends at the critical point when the waterfall field rolls rapidly to its minimum acquiring a non-zero symmetry breaking vev. In this case tachyonic, not parametric, preheating dominates and rapidly changes  the vacuum energy into topological structures involving the waterfall field \footnote{This is the conclusion of  \cite{Felder:2000hj}, \cite{Felder:2001kt} for a similar waterfall field case.}. One still has to convert this energy to SM states and this happens through normal perturbative reheating  \cite{Desroche:2005yt}. The inflaton field also rolls to its minimum with $<\phi>=\pi f$ and, in contrast to natural inflation, acquires an additional contribution to its mass, $\delta m_\phi=\sqrt{\delta/2h_1}\;m_\chi$, from its coupling to the waterfall field.  

Both fields can couple to the SM Higgs, $h$, via the couplings allowed by the symmetries of the model, $k_\phi (\Phi^2+\Phi^{\dagger 2})h^2\supset k_\phi\phi^2 h^2$ and $k_\chi \chi^2 h^2$. 
However the reheating temperature is strongly constrained by the fact the couplings $k_{\phi,\chi}$ must be small enough not to generate an unacceptably large mass for the Higgs. Taking account of this the most important couplings for reheating to the SM fields are to the top quark and are of the form $k'_\phi (\Phi^2+\Phi^{\dagger 2})m_t \bar t t/M^2\supset k'_\phi\phi^2 m_t \bar t t/M^2$ and $k'_\chi \chi^2 h^2m_t \bar t t/M^2$ where $M$ is a mediator mass coming from the UV sector of the theory above the cut-off scale $\Lambda$. The maximum possible value of the couplings corresponds to $M^2\sim<\phi>^2$, $<\chi>^2$ giving $k'_{\phi,\chi}\le O(1)$. For the case the other mass scales in the theory are close, $\phi\sim\chi\sim\Delta$, the couplings $\phi \bar t t$ and $\chi \bar t t$ governing the decay rate to top quarks are suppressed by a factor $x\le m_t/\Delta$. For this case it is easy to determine the reheat temperature from the condition $\Gamma_{\phi,\chi}>H(T_{rh})$. Before decay the  fields  are non-relativistic so  $H(T)^2\sim{\Delta^4\over M_P^2}\left({T\over\Delta}\right)^3$ giving $T_{rh}\sim m_t\left({M_P^2m_t\over\Delta^3}\right)^{1/3}$. If we require that the reheat temperature should be above the electroweak scale there is an upper limit on the inflation scale given by $\Delta<10^{13}GeV$. 

Clearly this conclusion follows because the SM states to which the inflaton and waterfall fields decay are light. The bound can be evaded if the principle decay is to heavy states. An obvious possibility is that the decay is to heavy right-handed SM singlet neutrinos, $\nu_R$, that allow for small neutrino masses through the see-saw mechanism. If these states are present the decay rate to them is enhanced over the decay to top quarks by the factor $(m_{\nu_R}/m_t)^2$, provided that $m_{\nu_R}<\Delta$ so that the decay can proceed. In this case  it is clear that heating is efficient with $T_{rh}\sim\Delta$.
\subsubsection{Model 2}
The enlarged $Z_2\times Z_2'$ symmetry of the second model restricts the allowed coupling between the fields $\phi$ and $\chi$ and SM states.  The relevant coupling determining the reheat temperature is that to the top quark  and is given by the terms $(\Phi^4+\Phi^{\dagger 4})m_t \bar t t/M^4$ and $(\chi^2+\chi^{\dagger 2})m_t \bar t t/M^2$. However if we allow for the minimum possible mediator scale, as we did above, the suppression, $x$, of these couplings is the same as before so the bounds on the reheat temperature are given as before. For the case  $\Delta<10^7 GeV$ the reheating is efficient with $T_{rh}\sim\Delta$. For larger $\Delta$, if we require that the reheat temperature should be above the electroweak scale, there is an upper limit on the inflation scale given by $\Delta<10^{13}GeV$. 
Allowing for the decay to $\nu_R$ these bounds are evaded and efficient reheating is possible over the full range $\Delta\le 10^{16} GeV$.

\section{Slow-roll parameters and observables} \label{Slow} 
Having shown how the EFT HNI potential can result from simple models we turn to a discussion of the inflationary predictions of the model. Before doing so, however, we gather a set of formulas for the SR parameters and observables of the model which are discussed in the rest of the paper. We also write expressions for the number of e-folds $N$ which are useful for later sections.

The inflationary sector of HNI is given by the potential Eq.(\ref{pot1}). In the slow-roll approximation the spectral indices are given in terms of the SR parameters of the model which involve the potential $V(\phi)$ and its derivatives (see e.g. Liddle:1994dx}, \cite{Liddle:2000cg})
\begin{equation}
\epsilon \equiv \frac{M^{2}}{2}\left( \frac{V^{\prime }}{V }\right) ^{2},\quad
\eta \equiv M^{2}\frac{V^{\prime \prime }}{V}, \quad
\xi_2 \equiv M^{4}\frac{V^{\prime }V^{\prime \prime \prime }}{V^{2}},\quad
\xi_3 \equiv M^{6}\frac{V^{\prime 2 }V^{\prime \prime \prime \prime }}{V^{3}}.
\label{Slowparameters}
\end{equation}%
Here primes denote derivatives with respect to the inflaton $\phi$ and $M = 2.44 \times 10^{18} GeV$ is the
reduced Planck mass which, for convenience, we set $M=1$. Defining $c_{\phi}$ and $s_{\phi}$ by $\cos(\frac{\phi}{f})$ and $\sin(\frac{\phi}{f})$ respectively, we get
\begin{eqnarray}
\epsilon &=&\frac{1}{2}\left(\frac{a}{f}\right)^2\frac{s_{\phi} ^{2}}{\left( 1+a\, c_{\phi} \right)^2} ,
\label{HNIeps}%
\\
\eta &=&-\left( \frac{a}{f^2}\right)\, \frac{c_{\phi}}{1+a c_{\phi}} , 
\label{HNIeta1}
\\
\xi_2 &=&-\left( \frac{a}{f^2}\right)^2\,\frac{1-c_{\phi} ^{2}}{\left( 1+a\, c_{\phi} \right)^2} ,\\
\xi_3 &=& +\left( \frac{a}{f^2}\right)^3\,\frac{1-c_{\phi} ^{2}}{\left(1+a c_{\phi}\right)^3} c_{\phi}.
\label{HNIzeta}
\end{eqnarray}%

In the SR approximation observables are given by (see e.g., \cite{Liddle:2000cg})
\begin{eqnarray}
n_{t} &=&-2\epsilon =-\frac{r}{8} , \label{Int} \\
n_{s} &=&1+2\eta -6\epsilon ,  \label{Ins} \\
n_{sk} &\equiv&\frac{d n_{s}}{d \ln k}=16\epsilon \eta -24\epsilon ^{2}-2\xi_2, \label{Insk} \\
n_{skk} &\equiv&\frac{d^{2} n_{s}}{d \ln k^{2}}=-192\epsilon ^{3}+192\epsilon ^{2}\eta-
32\epsilon \eta^{2} -24\epsilon\xi_2 +2\eta\xi_2 +2\xi_3, \label{Inskk} \\
A_s(k_H) &=&\frac{1}{24\pi ^{2}}\frac{\Delta^4}{%
\epsilon _H},
\label{IA} 
\end{eqnarray}
where  $n_{sk}$ denotes the running of the scalar index and $n_{skk}$ the running of the running, in a 
self-explanatory notation. The density perturbation at wave number $k$ is $A_s(k)$ with amplitude at horizon crossing given by $A_s(k_H)\approx 2.2\times 10^{-9}$ \cite{Ade:2015xua}.
The scale of inflation is 
$\Delta$ with $\Delta \approx V_{H}^{1/4}$ and $r\equiv A_t/A_s$ the ratio of tensor to scalar perturbations. 
All quantities with a subindex ${}_H$ are evaluated at the scale $\phi_{H}$, at which observable perturbations 
are produced, some $50-60$ e-folds before the end of inflation. 

We can now express the number of e-folds from $\phi_{H}$ to the end of inflation at $\phi _{e}$ as follows
\begin{equation}
N\equiv -\int_{\phi _H}^{\phi_e}\frac{V({\phi })}{V^{\prime }({\phi })}{d}{\phi }=\frac{f^2}{2a}\left((1+a)\ln \left( \frac{1-c_e}{1-c_H}\right)+(1-a)\ln \left( \frac{1+c_H}{1+c_e}\right)\right),
\label{N1}
\end{equation}
where $c_{H}\equiv \cos \left( \frac{\phi _{H}}{f}\right)$ and $c_{e}\equiv \cos \left( \frac{\phi _{e}}{f}\right)$. Eq. (\ref{N1}) can also be written as
\begin{equation}
N=\frac{f^2}{a}\left[\ln\left(\tan(\frac{\phi}{2 f})\right)+a \ln\left(\sin(\frac{\phi}{f})\right)\right]\Bigl|_{\phi_H}^{\phi_e},
\label{N2}
\end{equation}
which will be particularly convenient in subsection \ref{approx}.
\section{HNI phenomenology}  \label{pheno} 
Here we discuss in detail the phenomenological implications of HNI, comparing them to the most recent Planck results. There are two regions of parameter space that require different treatments depending on whether  ${f\over M}$ is small or not. If it is, one can obtain accurate analytical results for the observables; if not, it is necessary to perform a numerical study. We consider these two cases in turn:

\subsection{Approximate analytic solution} \label{approx} 
During inflation the SR parameters $\epsilon$ and $\eta$ should satisfy $\epsilon << 1$ and $\eta << 1$. If ${f\over M}<<1$ this requires that $a<<1$  and so we may expand Eqs. (\ref{HNIeps}-\ref{HNIzeta}) as a power series in $a$ giving
\begin{eqnarray}
\epsilon_H &\approx&\frac{1}{2}\left(\frac{a}{f}\right)^2s_{H}^2 ,
\label{HNIepsapp}%
\\
\eta_H &\approx& -\left(\frac{a}{f^2}\right)c_{H} , 
\label{HNIetaapp}
\\
\xi_2 &=&-2\left( \frac{1}{f}\right)^2\,\epsilon_H ,\\
\xi_3 &=& \xi_2\eta_H.
\label{HNIxi3app}%
\end{eqnarray}%
The number of observables possibly measureable
 are $A_s(k),\;n_s,\;r,\;n_t,\;n_{sk},\;n_{skk}$ and $N$. The parameters of the effective field theory description of HNI models discussed are $\Delta,\;a,\;\phi_H,\;f$ and $\phi_e$ that can conveniently be replaced by $\Delta,\;\epsilon_H,\;\eta_H,\;f$ and $\phi_e$. Thus HNI gives are two relations between the observables. One follows immediately from the slow-roll conditions and is given by \cite{Liddle:2000cg}
 \be
 n_{t} =-2\epsilon_H =-\frac{r}{8}.
 \ee
 To determine the second relation note that, since  $\epsilon<<\eta$, we have
\begin{eqnarray}
\delta_{ns}\equiv 1-n_s&\approx& -2\eta_H,
\label{HNIeta}
\\
\xi_2 &\approx&-\frac{n_{sk}}{2}.
\end{eqnarray}%
Combining these two gives the second relation between observables
\be
n_{skk} \approx 4\eta\xi_2 =\delta_{ns} n_{sk}.
\label{nskk}
\ee
The remaining observables are then given in terms of the parameters by
\bea
r&=&16\epsilon_H,\nonumber\\
n_s&=&1+2\eta_H,\nonumber\\
n_{sk}&=&{r\over 4 f^2}.
\label{nsk}
\eea
Note that $n_{sk}$ and $n_{skk}$ are positive.
The remaining parameter, $\phi_e$, determines the number of-folds through Eq. (\ref{N2}) which in this region gives 
\be
N\approx {f^2\over a}\left[\ln\left(\tan({\phi_e\over 2f})\right)-\ln\left(\tan({\phi_H\over 2f})\right)\right],
\label{N}
\ee
where 
\be
\cos(\phi_H)^2={\eta_H^2\over({2\epsilon_H\over f^2}+\eta_H^2)}.
\ee
\subsubsection{The upper bound of r}\label{bound}
The requirement that uncorrelated parameters should not be taken to be arbitrarily close in magnitude leads to a "fine-tuning" bound on $r$. To see how this works consider eq(\ref{N}) for $N$.
As we discussed above the ``fine tuning'' constraint leads to a bound on how closely $\phi_e/f$ can approach $\pi$. At the 10\% level this translates to a bound $\ln(\tan(\phi_e/2f))\le1.5$ while at the 1\% level $\ln(\tan(\phi_e/2f))\le2.6$. As a result the constraint $N=60$ implies $f^2/a=$  $40$ and $23$ respectively.
Then from Eq. (\ref{HNIetaapp}) we find $r=16\epsilon_H<5\times10^{-3}f^2/M^2$ at the 10\% fine tuning level and $r=16\epsilon_H<1.5\times10^{-2}f^2/M^2$ at the 99\% fine tuning level. For sub-Planckian values of f this gives bounds on $r$ far below the Planck limits.

\subsubsection{The inflationary scale}
The most significant difference between the two models presented above is the lower limit on the inflationary scale.  From Eq. (\ref{IA}) we see that $\epsilon_H\propto \Delta^4$ and so reducing $\Delta$ implies $\epsilon$ becomes negligible in determining $\delta_{ns}$. 

For the first waterfall field model the fine-tuning constraint requires $\sin(\frac{\phi_H}{f})=O(1)$. In this case  the only way $\delta_{ns}$ can be consistent with the measured value is if $({a\over f})^2\propto \Delta^4$ with ${a\over f^2}$ constant. As a result we require $a\propto \Delta^4$ and $f\propto \Delta^2$. For small $\Delta$ the latter condition is inconsistent because, c.f. Eq. (\ref{f2}), $\tilde f\propto m_\phi$ which is not protected by a symmetry and so is of $O(\sqrt{\alpha} \Lambda)$ in the absence of fine tuning. The best one can do is in a low-scale completion where $\Lambda =O(1TeV)$ and in this case $\Delta\sim \sqrt(\alpha)(1 TeV)M\sim 10^{11}GeV$.

For the case of the second model the fine tuning constraint is consistent with small $\sin(\frac{\phi_H}{f})$ and so it is possible for $\epsilon$ to be very small through the smallness of ${\phi_H\over f}$. For small ${\phi_H\over f}$ Eqs. (\ref{HNIeps},\ref{HNIeta},\ref{Ins}) and (\ref{IA}) imply
\be
\Delta=M\sqrt{{A_H\delta_{ns}\over 2}({\phi_H\over M})}\approx 9\times 10^{15}\sqrt{{\phi_H\over M}}GeV.
\label{lowscale}
\ee
In this case $\Delta$ can be very small without limiting $f$ and so a very low scale of inflation, even down to the electroweak scale, is in principle possible. However the cosmological constant condition requires $\Delta\sim m_\chi \sim \sqrt{\alpha}\Lambda$ so the scale of UV completion cannot be much larger than the inflation scale.  
\subsubsection{Comparison of the analytic solution with the Planck 2015 data}
The most recent Planck analysis of inflationary models has produced an accurate measurement of $n_s$, an upper bound on $r$, and improved measurement of $A_s(k)$. In addition it has performed fits that provide limits on $n_{sk}$ and $n_{skk}$. Given this it is of interest to compare Planck's results with HNI. 

The HNI parameters $\eta_H$ and $\phi_e$ can be chosen to fit the observed value of $n_s$ and $N$. 
As mentioned above $n_{sk}$ and $n_{skk}$ are positive in HNI. The Planck fit, including $n_{sk}$ only,  indicates it is small with negative central value but consistent with zero at $1\sigma$. At $3\sigma$ we have $n_{sk}<0.016$ which, from Eq. (\ref{nsk}), requires $r<0.064$, for sub-Planckian values of $f$, consistent with the Planck bound $r<0.11$.  From Eq. (\ref{IA}) and the measured value of $A_s(k)$, this limit on  $r$ implies $\Delta<4\times 10^{-16}GeV$. These  limits become much stronger for smaller values of $f/M$. 
When both $n_{sk}$ and $n_{skk}$ are allowed the Planck fit  gives positive central values for $n_{sk}=0.011\pm0.014$ and $n_{skk}=0.029\pm0.016$, with $n_s=0.9569\pm0.0077$. Assuming the central Planck value for $n_{sk}$ the corresponding HNI prediction from Eq. (\ref{nskk}) is $n_{skk}=4.7\times 10^{-4}$, consistent at $2\;\sigma$ with the Planck fit.

Overall it is clear that an excellent fit to the data is possible in the sub-Planckian $f$ region but, due to the number of parameters of the model, the data does not provide a stringent test of the model. For the case of model 1 we have $c_H\approx 1$ so the limit on $r$ provides the limit ${a\over f}<0.089$ but the data on $n_s$ cannot be used to determine $a$ and $f$ separately because $c_H$ is not well determined. For the case of model 2 we have $s_H\approx 1$ so the measurement of $n_s$ gives ${a\over f^2}\approx 0.02$, but then the limit on $r$ cannot be used to determine $a$ and $f$ separately because now $s_H$ is not well determined.
\subsection{Detailed numerical fit of HNI to the available inflationary  data }\label{num}

Due to the correlation of the HNI predictions for the inflationary observables\footnote{For example we noted above that $n_{sk}$ and $n_{skk}$ are positive.} it is necessary to perform a numerical fit to all available data in order to determine the range of observables consistent with HNI and hence to map out the significant tests of the model.

To carry out the exploration of the parameter space, we incorporated the 
predictions of HNI in the standard
cosmological equations by performing minor modifications to the CAMB code \cite{Lewis:1999bs}. We then include it in the CosmoMC software \cite{Lewis:2002ah}  and this was used to fit all available data. In particular, we provide constraints on HNI  by using the temperature (TT) and polarization (low P) measurements from the 2015 data release of the Planck experiment along with the B-mode polarization constraints from a joint analysis 
of BICEP2, Keck Array, and Planck (BKP) \cite{Ade:2015tva}.
Throughout the analysis we consider purely Gaussian adiabatic perturbations and, at the background level,
assume the standard $\Lambda$CDM model specified by the following sampling parameters: 
the physical
baryon density $\Omega_{\rm b} h^2$ and CDM density $\Omega_{\rm DM}
h^2$, where $h$ is the dimensionless Hubble parameter such that
$H_0=100h$ kms$^{-1}$Mpc$^{-1}$; $\theta$, which is $100\times$ the ratio of the sound
horizon to angular diameter distance at last scattering surface; the
optical depth $\tau$ at reionisation; and parameters describing the primordial power spectra: the amplitude $A_{\rm s}$
of the primordial perturbation spectrum, the scale parameter $f$, $a$,
 the inflaton field when cosmological scales leave the horizon $\phi_H$, 
and the tensor-to-scalar ratio $r$.
The ranges of the uniform flat
priors assumed on these standard LCDM parameters are the following: $\Omega_{\rm b} h^2 = [0.01,0.03]$, 
$\Omega_{\rm DM}h^2 = [0.05, 0.20]$, $\theta = [1, 1.1]$,  $\tau=[0.01,0.3]$, $\ln[10^{10}A_{\rm s}] = [2.5, 4]$; 
and two conservative  cases (1) $\ln f =[-5,0]$, $\ln a = [-12,0]$, $\ln \phi_{\rm H} =[-5,\pi f]$, (2) $f = [0,6]$, $\ln a = [-4, 0]$, $\phi_{\rm H} =[0,\pi f]$.

\begin{figure}[h!]
\begin{center}
\includegraphics[trim = 2mm 1mm 1mm 1mm, clip, width=18cm, height=6cm]{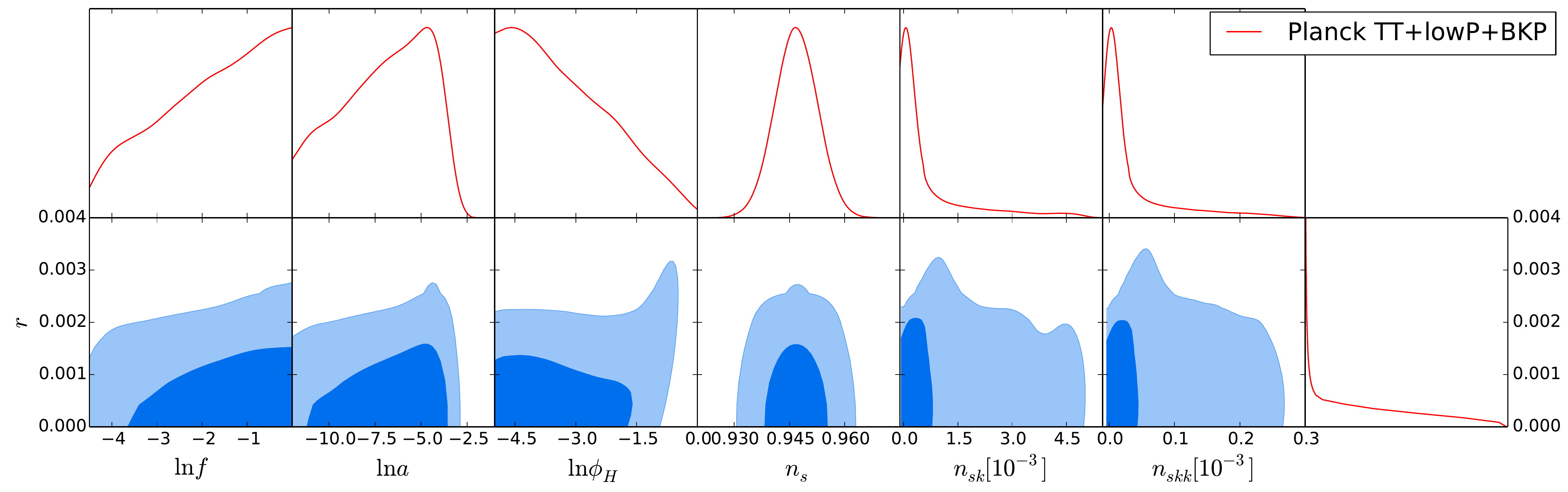}
\caption{1D and 2D marginalised posterior distributions on density parameters of the HNI model for CMB {\it{Planck}-TT 2015 data, Polarization information (low P) and the B-mode polarization constraints from a joint analysis of BICEP2, Keck Array, and Planck (BKP) data.} Note that $n_{sk}$ and $n_{skk}$ are always positive, this is determined by the fact that the symmetry breaking scale $f$ takes sub-Planckian values. Comparison with Fig. (22) of \cite{Ade:2015xua} shows that all values of $r$ and $ns$ above are contained by the black contours of the $\Lambda$CDM + running + tensors model using Planck data.}
\label{planck}
\end{center}
\end{figure}

\begin{figure}[h!]
\begin{center}
\includegraphics[trim = 2mm 1mm 1mm 1mm, clip, width=18cm, height=6cm]{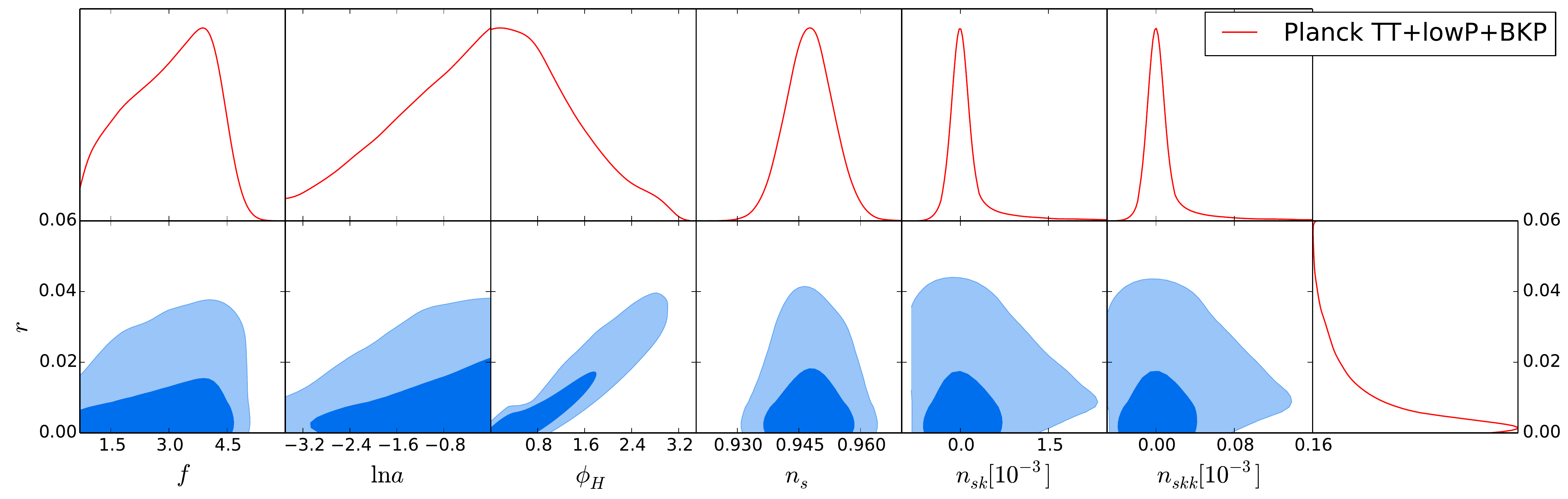}
\caption{Same as Fig. \ref{planck} but with $0 < f$. Negative values of $n_{sk}$ and $n_{skk}$ in HNI require super-Planckian values of $f$ \cite{German:2015qjq}.}
\label{planck2}
\end{center}
\end{figure}

For the case of sub-Plankian values of $f$, Fig.~\ref{planck} displays 1D and 2D marginalised posterior distributions on density parameters of the HNI model. 
The observables describing the running of the  scalar power spectrum $n_{sk}$ and $n_{skk}$ respectively, satisfy the following relations 
\cite{Carrillo-Gonzalez:2014tia}
\begin{eqnarray}
n_{sk}&=&\frac{r}{32}\left(3r-16\delta_{ns}+\frac{8}{f^2} \right),
\label{HNInsk}\\
n_{skk}&=&\frac{r}{128}\left( 3r^2+\frac{12}{f^2}r-32\left( 2\delta_{ns}-\frac{1}{f^2} \right)\delta_{ns}  \right).
\label{HNInskk}
\end{eqnarray}
From here one can see that a scale parameter $f<1$ does not allow for the possibility of a negative $n_{sk}$ or $n_{skk}$. Thus, a detection of a negative running would require the scale $f$ to be super-Planckian as occurs in natural inflation \cite{German:2015qjq}. To allow for this we plot in 
Fig.~\ref{planck2} the same quantities as in Fig.~\ref{planck}, dropping the constraint on $f$. Table \ref{tab:results} gives the corresponding constraints of the fit for the two cases.

\begin{table*}[htbp]
\begin{center}
\scalebox{0.9}{
\begin{tabular}{ccccccc}
\hline
 $f$  & $\ln a$	    & 	$\phi_{\rm H}$ &  $n_s$& \quad $n_{sk}[10^{-3}]$  & \quad$n_{skk}[10^{-3}]$ &\quad $r$\\
\hline
\hline
\\
 unconstrained & $< -3.24$  &$\ln \phi_{\rm H}$ $<- 1.1$	&	\quad$0.9467 \pm0.0055$  \quad  & $<3.1$  & $<0.16$ &\quad$<0.017$ \\
\hline
\\
  $< 4.67$ & unconstrained & $< 2.5$	&$0.947 \pm  0.005$ &$ 0.17\pm  0.59$ &  $0.011\pm  0.036$ & $< 0.03$ \\
\hline
\end{tabular}
}
\bigskip
\caption{Constraints on HNI parameters. In the first row $0 < a < 1$ and $0 < f < 1$, in the second $0 < a < 1$ and $f > 0$. For one-tailed distributions the upper limit 95\% CL is given. For two-tailed the 68\% is shown.}
\label{tab:results}
\end{center}
\end{table*}

\section{Abundance of primordial black hole production and hierarchy of slow-roll parameters} \label{pbh}
At first order in the SR parameters the scalar power spectrum is given by
\begin{equation}
\mathcal{P}_s(k)=\frac{1}{24\pi^2M^4}\frac{V}{\epsilon}=A_s \left( \frac{k}{k_H}\right)^{(n_s-1) + \frac{1}{2}n_{sk}  \ln\left(\frac{k}{k_H}\right) +\, \cdot \, \cdot \, \cdot  }.\label{power}
\end{equation}
It has been shown \cite{Kohri:2007qn} that  there exists an additional constraint coming from the possible over-production of primordial black holes (PBHs) at the end of inflation. 
Due to this constraint the Taylor expansion of the power spectrum around its  value at horizon crossing, $N_{H}\approx 60$ is bounded by
\footnote{To lowest order in slow-roll $d/dN =- d / d\ln k$. The next order term in the expansion of Eq.~\eqref{ps:expansion}, involving the parameter $n_{skk}$, is subdominant.},
\begin{equation}
\label{ps:expansion}
 \ln \left[\frac{\mathcal{P}_s(0)}{\mathcal{P}_s(N_H)}\right] = (n_s- 1) N_H + \frac{1}{2} n_{sk} N_H^2\leq 14,
\end{equation}
where $\mathcal{P}_s(N= 0)\simeq 10^{-3}$ (see also Refs.~\cite{Josan:2009qn,Carr:2009jm}) evolves from the initial value $\mathcal{P}_s(N_H) \approx 10^{-9}$. 
This gives the bound $n_{sk}< 10^{-2}$ that is readily satisfied by HNI ({\it c.f.} Table \ref{tab:results}).
Note that the validity of the approximation, Eq. (\ref{ps:expansion}), requires an hierarchy of SR parameters to be satisfied \cite{Kohri:2007qn} , i.e., if $\epsilon<<\eta$ the hierarchy of SR parameters required is
$ \Bigl|\xi_{m+1}\Bigl|<<\Bigl|\xi_{m}\Bigl|$ where (renaming the SR parameter $\eta$ by $\xi_1$)
\begin{equation}
\xi_1 \equiv M^{2}\frac{V^{\prime \prime }}{V}, \quad
\xi_2 \equiv M^{4}\frac{V^{\prime }V^{\prime \prime \prime }}{V^{2}},\quad
\xi_3 \equiv M^{6}\frac{V^{\prime 2 }V^{\prime \prime \prime \prime }}{V^{3}}, \quad
\xi_4 \equiv M^{8}\frac{V^{\prime 3 }V^{\prime \prime \prime \prime \prime}}{V^{4}}\cdot \cdot \cdot .
\label{SRP}
\end{equation}%
For the HNI potential these can be written as
\begin{eqnarray}
\xi_1 &=&-\left( \frac{a}{f^2}\right)\, \frac{c_{\mathrm{\phi}}}{1+a c_{\mathrm{\phi}}} , \\
\xi_2 &=&-\left( \frac{a}{f^2}\right)^2\,\frac{s_{\phi}^2}{\left( 1+a\, c_{\mathrm{\phi}} \right)^2} ,\\
\xi_3 &=& +\left( \frac{a}{f^2}\right)^3\,\frac{c_{\mathrm{\phi}} s_{\phi}^2}{\left(1+a c_{\mathrm{\phi}}\right)^3} =\xi_1\xi_2 ,\\
\xi_4 &=& \xi_2^2 , \quad \quad \xi_5=\xi_1\xi_2^2, \quad \quad \xi_6=\xi_2^3, \cdot \cdot \cdot
\end{eqnarray}%
In HNI at $\phi_H$, $\epsilon<<\eta<<1$  and $\Bigl|\xi_2\Bigl|<<\Bigl|\xi_1\Bigl|$ follows from the fact that $\Bigl|\xi_2\Bigl|=2\epsilon/f^2 \sim (a/f^2)^2 s_{\phi}^2\sim \eta^2 s_{\phi}^2<<\Bigl|\eta\Bigl|=\Bigl|\xi_1\Bigl|$.
From here we see that $\xi_3=\xi_1\xi_2<<\xi_2$, $\xi_4=\xi_2^2=\xi_2\xi_2<<\xi_1\xi_2=\xi_3$, $\xi_5=\xi_1\xi_2^2<<\xi_2^2=\xi_4$, and so on. Thus the required hierarchy of SR parameters is guaranteed in HNI.

Further PBH production can occur when the roll of the waterfall field is ``mild", in the sense that there is an appreciable number of e-folds of inflation generated after the waterfall field starts to roll \cite{Clesse:2010iz}-\cite{Mulryne:2011ni}.  When large curvature perturbations are generated at the end of the {\it valley phase} of inflation i.e., after the inflaton has reached the critical point, $\phi_c$, defined as the point where the waterfall starts, PBH are produced and it is important to determine the constraints on the  model parameters so that the PBH production does not conflict with CMB constraints on its abundances. Particularly interesting is the suggestion that PBH might be dark matter candidates\cite{Carr:1974nx}-\cite{Blais:2002nd} and this certainly deserves further study. A first analysis of this possibility in a hybrid inflation model suggested it might indeed  produce PBH dark matter and act as seeds of galaxies \cite{Clesse:2015wea}. However a more recent non-perturbative numerical study \cite{Kawasaki:2015ppx} of the curvature perturbation produced during the waterfall phase concluded that if there are more than 5 e-folds of inflation during the waterfall stage, there will be an unacceptable rate of black hole production. Moreover the mass scale of the produced PBH is of  $ \mathcal{O}(10)kg\left(\frac{H_{inf}}{10^9 GeV}\right)^{-1}$  which evaporate soon after production and so cannot make up dark matter.

 In the hybrid models discussed here it is relatively easy to limit the number of e-folds during the waterfall phase.
In the first model the condition that there should be no more than 5 e-folds of ${\it waterfall}$ inflation follows from the constraint on the $\Pi$ parameter \cite{Clesse:2015wea,Kawasaki:2015ppx} that, to a good approximation, determines the amount of waterfall inflation. Applying this constraint we find
\begin{equation}
\Pi^{-2}=\left(\frac{M_p}{\Delta}\right)^4 a \delta \sin^2\left(\frac{\phi_c}{f}\right)> \frac{1}{10}\,.
\end{equation}
Using Eq. (\ref{HNIeta}) to eliminate $a$ we get
\begin{equation}
x>20\frac{\cos(\phi_H/f)}{\sin^2(\phi_c/f)},
\end{equation}
where $4\delta f^2=x(\Delta^4/M^2_P)$, (c.f. Eq. (\ref{massconstraint})). The constraint on fine tuning discussed above limits how small the denominator can be and at the $10\%$ level shows that  $x>100$ is sufficient  condition to keep waterfall inflation at an acceptable level, although there are regions of parameter space where $x$ can be much closer to the original constraint, $x>1$, following from Eq. (\ref{massconstraint}).

In the second model the constraint is given by
\begin{equation}
x>20\frac{\cos(\phi_H/f)}{\sin(\phi_c/f)\cos(\phi_c/2f)}.
\end{equation}
In this model it is possible for the sine term to be very small, corresponding to the low inflation scale limit, so it is important to examine this limit in detail. Imposing a slightly stronger limit on $x$ we can guarantee the desired condition by 
\begin{equation}
x>\frac{20}{\sin(\phi_H/f)}>\frac{20}{\sin(\phi_c/f)}. 
\end{equation}
Now from Eq. (\ref{lowscale}) we have 
\begin{equation}
\sin(\frac{\phi_H}{f})\approx \frac{\phi_H}{f} \approx  \left(\frac{M}{9\times 10^{15}}\right)^2 \left(\frac{\Delta}{M}\right)^2\frac{M}{f} \approx 7\times 10^4 \left(\frac{\Delta}{M}\right)^2\frac{M}{f},
\end{equation}
so $x>\frac{20}{7\times 10^4} \left(\frac{M}{\Delta}\right)^2\frac{f}{M}$. Thus $4 \delta f^2=x\left(\frac{\Delta^4}{M^2}\right)>3\times 10^{-4}\frac{ \Delta^2}{M}f$, i.e., 
\begin{equation}
4 \delta f>3\times 10^{-4}\frac{\Delta^2 }{M}. 
\label{limit}
\end{equation}
Finally we can combine this with the bound of Eq. (2.22) in the form $2\sqrt{2}\delta^{1/2}f>\Delta^2/M$ to obtain 
\begin{equation}
4 \delta f=\sqrt{2}\delta^{1/2} 2\sqrt{2}\delta^{1/2}f>\sqrt{2}\delta^{1/2}\Delta^2/M. 
\end{equation}
Thus the constraint of eq(\ref{limit}) is satisfied if $\delta^{1/2}>\frac{3}{\sqrt{2}}\times 10^{-4}\approx 2\times 10^{-4}$. In conclusion the low inflation scale limit does not lead to an overproduction of PBH if the coupling obeys the mild limit $\delta>4\times 10^{-8}$. 

\section{Summary and conclusions} \label{conclu} 
We have shown that it is straightforward to construct hybrid versions of Natural Inflation in which a waterfall field coupled to the Pseudo-Goldstone inflaton is responsible for ending inflation. The models require a discrete symmetry to order the breaking of the underlying continuous symmetry responsible for the mass of the would-be Goldstone mode. Two models were constructed that demonstrate the range of possibilities, one with an extended discrete symmetry allowing for very low scales of inflation.

In contrast to the original Natural Inflation model the hybrid models allow for an acceptable inflationary era even with a sub-Planckian spontaneous breaking of the Goldstone symmetry.  For the case that reheating proceeds through the coupling of the inflaton and waterfall field to SM states there is an upper bound on the reheat temperature that in turn provides a significant upper bound on the inflationary scale. This bound can be evaded if the decay is to heavy states, such as heavy right-handed neutrinos.

In Hybrid Natural Inflation the slow-roll parameter $\epsilon$ is a non-monotonic function of the inflaton  field with a maximum where observables take universal  values that determines the maximum possible tensor to scalar ratio, $r$. A detailed analytic study of the model was presented and compared to the Planck 2015 temperature and polarisation data, showing excellent agreement for a wide range of the underlying parameters and inflationary scale and satisfying the  constraints coming from non-overproduction of Primordial Black Holes. A full numerical fit to all available inflationary data was also presented, establishing the possible range of observables consistent with HNI and thus mapping out the possible tests of the model.

\vspace{0.7cm}
\noindent{\bf \Large Acknowledgements}

\vspace{0.5 cm}

G.G. gratefully acknowledges financial support from PASPA-DGAPA, UNAM and CONACYT, Mexico and the hospitality of the Rudolf Peierls Centre for Theoretical Physics, Oxford where this work has been carried out. He also acknowledges  support from \textit{Programa de Apoyo a Proyectos de Investigaci\'on e Innovaci\'on Tecnol\'ogica} (PAPIIT) UNAM, IN103413-3, \textit{Teor\'ias de Kaluza-Klein, inflaci\'on y perturbaciones gravitacionales}. We are also grateful to Subir Sarkar for useful comments.


\begin{thebibliography}{99}  

\bibitem{Guth:1980zm}
  A.~H.~Guth,
  Phys.\ Rev.\ D {\bf 23} (1981) 347.
  doi:10.1103/PhysRevD.23.347

\bibitem{Linde:1981mu}
  A.~D.~Linde,
  Phys.\ Lett.\ B {\bf 108} (1982) 389.
  doi:10.1016/0370-2693(82)91219-9

\bibitem{Albrecht:1982wi}
  A.~Albrecht and P.~J.~Steinhardt,
  Phys.\ Rev.\ Lett.\  {\bf 48} (1982) 1220.
  doi:10.1103/PhysRevLett.48.1220

\bibitem{Lyth:1998xn}
  D.~H.~Lyth and A.~Riotto,
  Phys.\ Rept.\  {\bf 314} (1999) 1
  doi:10.1016/S0370-1573(98)00128-8
  [hep-ph/9807278].
  D.~H.~Lyth and A.~R.~Liddle,
  {\it The primordial density perturbation: Cosmology, inflation and the origin of structure},
  Cambridge, UK: Cambridge Univ. Pr. (2009) 497 p;
D.~Baumann,
  arXiv:0907.5424 [hep-th].

\bibitem{Freese:1990rb}
  K.~Freese, J.~A.~Frieman and A.~V.~Olinto,
  Phys.\ Rev.\ Lett.\  {\bf 65} (1990) 3233.
  doi:10.1103/PhysRevLett.65.3233

\bibitem{Adams:1992bn}
  F.~C.~Adams, J.~R.~Bond, K.~Freese, J.~A.~Frieman and A.~V.~Olinto,
  Phys.\ Rev.\ D {\bf 47} (1993) 426
  doi:10.1103/PhysRevD.47.426
  [hep-ph/9207245].
  
\bibitem{Freese:2008if}
  K.~Freese, C.~Savage and W.~H.~Kinney,
  Int.\ J.\ Mod.\ Phys.\ D {\bf 16} (2008) 2573
  doi:10.1142/S0218271807011371
  [arXiv:0802.0227 [hep-ph]].

\bibitem{Freese:2014nla}
  K.~Freese and W.~H.~Kinney,
  JCAP {\bf 1503} (2015) 044
  doi:10.1088/1475-7516/2015/03/044
  [arXiv:1403.5277 [astro-ph.CO]].
  
\bibitem{Planck:2013jfk}
  P.~A.~R.~Ade {\it et al.} [Planck Collaboration],
  Astron.\ Astrophys.\  {\bf 571} (2014) A22
  doi:10.1051/0004-6361/201321569
  [arXiv:1303.5082 [astro-ph.CO]].

\bibitem{Kim:2004rp}
  J.~E.~Kim, H.~P.~Nilles and M.~Peloso,
  JCAP {\bf 0501} (2005) 005
  doi:10.1088/1475-7516/2005/01/005
  [hep-ph/0409138].


\bibitem{Dimopoulos:2005ac}
  S.~Dimopoulos, S.~Kachru, J.~McGreevy and J.~G.~Wacker,
  JCAP {\bf 0808} (2008) 003
  doi:10.1088/1475-7516/2008/08/003
  [hep-th/0507205].
  
\bibitem{Palti:2015xra}
  E.~Palti,
  JHEP {\bf 1510} (2015) 188
  doi:10.1007/JHEP10(2015)188
  [arXiv:1508.00009 [hep-th]].
  
\bibitem{Baume:2016psm}
  F.~Baume and E.~Palti,
  arXiv:1602.06517 [hep-th].
  

\bibitem{Ross:2009hg}
  G.~G.~Ross and G.~German,
  Phys.\ Lett.\ B {\bf 684} (2010) 199
  doi:10.1016/j.physletb.2010.01.003
  [arXiv:0902.4676 [hep-ph]].

\bibitem{Ross:2010fg}
  G.~G.~Ross and G.~German,
  Phys.\ Lett.\ B {\bf 691} (2010) 117
  doi:10.1016/j.physletb.2010.06.017
  [arXiv:1002.0029 [hep-ph]].






\bibitem{Vazquez:2014uca}
  J.~A.~Vazquez, M.~Carrillo-Gonzalez, G.~German, A.~Herrera-Aguilar and J.~C.~Hidalgo,
  JCAP {\bf 1502} (2015) no.02,  039
   Addendum: [JCAP {\bf 1510} (2015) no.10,  A01]
  doi:10.1088/1475-7516/2015/02/039, 10.1088/1475-7516/2015/10/A01
  [arXiv:1411.6616 [astro-ph.CO]].


\bibitem{Carrillo-Gonzalez:2014tia}
  M.~Carrillo-Gonzalez, G.~German, A.~Herrera-Aguilar, J.~C.~Hidalgo and R.~Sussman,
  Phys.\ Lett.\ B {\bf 734} (2014) 345
  doi:10.1016/j.physletb.2014.05.062
  [arXiv:1404.1122 [astro-ph.CO]].

\bibitem{Linde:1993cn}
  A.~D.~Linde,
  Phys.\ Rev.\ D {\bf 49} (1994) 748
  doi:10.1103/PhysRevD.49.748
  [astro-ph/9307002].
Other general references for hybrid inflation are: 
D.H. Lyth, A. Riotto, Phys. Rep. 314 (1999) 1, arXiv:hep-ph/9807278;
A.D. Linde, Lecture Notes in Phys. 738 (2008) 1, arXiv:0705.0164 [hep-th];
M. Shaposhnikov, J. Phys. Conf. Ser. 171 (2009) 012005.

\bibitem{East:2015ggf}
  W.~E.~East, M.~Kleban, A.~Linde and L.~Senatore,
  arXiv:1511.05143 [hep-th].

\bibitem{Hawking:1998bn}
  S.~W.~Hawking and N.~Turok,
  Phys.\ Lett.\ B {\bf 425} (1998) 25
  doi:10.1016/S0370-2693(98)00234-2
  [hep-th/9802030].
  
\bibitem{Vilenkin:1998pp}
  A.~Vilenkin,
  Phys.\ Rev.\ D {\bf 57} (1998) 7069
  doi:10.1103/PhysRevD.57.7069
  [hep-th/9803084].

\bibitem{Linde:1998gs}
  A.~D.~Linde,
  Phys.\ Rev.\ D {\bf 58} (1998) 083514
  doi:10.1103/PhysRevD.58.083514
  [gr-qc/9802038].
  
\bibitem{Turok:1998he}
  N.~Turok and S.~W.~Hawking,
  Phys.\ Lett.\ B {\bf 432} (1998) 271
  doi:10.1016/S0370-2693(98)00651-0
  [hep-th/9803156].

\bibitem{Turok:2000bt}
For a review, see, 
  N.~Turok,
  AIP Conf.\ Proc.\  {\bf 555} (2001) 14
  doi:10.1063/1.1363506
  [astro-ph/0011195].

\bibitem{Vilenkin:1982de}
  A.~Vilenkin,
  Phys.\ Lett.\ B {\bf 117} (1982) 25.
  doi:10.1016/0370-2693(82)90866-8
  
\bibitem{Vilenkin:1983xq}
  A.~Vilenkin,
  Phys.\ Rev.\ D {\bf 27} (1983) 2848.
  doi:10.1103/PhysRevD.27.2848
  
\bibitem{Vilenkin:1984wp}
  A.~Vilenkin,
  Phys.\ Rev.\ D {\bf 30} (1984) 509.
  doi:10.1103/PhysRevD.30.509
  
\bibitem{Linde:1983cm}
  A.~D.~Linde,
  Sov.\ Phys.\ JETP {\bf 60} (1984) 211
   [Zh.\ Eksp.\ Teor.\ Fiz.\  {\bf 87} (1984) 369].
  
\bibitem{Rubakov:1984ki}
  V.~A.~Rubakov,
  JETP Lett.\  {\bf 39} (1984) 107
   [Pisma Zh.\ Eksp.\ Teor.\ Fiz.\  {\bf 39} (1984) 89].
  
\bibitem{Zeldovich:1984vk}
  Y.~B.~Zeldovich and A.~A.~Starobinsky,
  Sov.\ Astron.\ Lett.\  {\bf 10} (1984) 135.
  
\bibitem{Vilenkin:1994pv}
  A.~Vilenkin,
  Phys.\ Rev.\ Lett.\  {\bf 72} (1994) 3137
  doi:10.1103/PhysRevLett.72.3137
  [hep-th/9402085].
  
\bibitem{Brandenberger:2016uzh}
  R.~Brandenberger,
  arXiv:1601.01918 [hep-th].

\bibitem{Kofman:1994rk}
  L.~Kofman, A.~D.~Linde and A.~A.~Starobinsky,
  Phys.\ Rev.\ Lett.\  {\bf 73} (1994) 3195
  doi:10.1103/PhysRevLett.73.3195
  [hep-th/9405187].

\bibitem{Shtanov:1994ce}
  Y.~Shtanov, J.~H.~Traschen and R.~H.~Brandenberger,
  Phys.\ Rev.\ D {\bf 51} (1995) 5438
  doi:10.1103/PhysRevD.51.5438
  [hep-ph/9407247].
  
\bibitem{Kofman:1997yn}
  L.~Kofman, A.~D.~Linde and A.~A.~Starobinsky,
  Phys.\ Rev.\ D {\bf 56} (1997) 3258
  doi:10.1103/PhysRevD.56.3258
  [hep-ph/9704452];
  
\bibitem{Felder:2000hj}
  G.~N.~Felder, J.~Garcia-Bellido, P.~B.~Greene, L.~Kofman, A.~D.~Linde and I.~Tkachev,
  Phys.\ Rev.\ Lett.\  {\bf 87} (2001) 011601
  doi:10.1103/PhysRevLett.87.011601
  [hep-ph/0012142].
  
\bibitem{Felder:2001kt}
  G.~N.~Felder, L.~Kofman and A.~D.~Linde,
  Phys.\ Rev.\ D {\bf 64} (2001) 123517
  doi:10.1103/PhysRevD.64.123517
  [hep-th/0106179].
  
\bibitem{Desroche:2005yt}
  M.~Desroche, G.~N.~Felder, J.~M.~Kratochvil and A.~D.~Linde,
  Phys.\ Rev.\ D {\bf 71} (2005) 103516
  doi:10.1103/PhysRevD.71.103516
  [hep-th/0501080].
  
  
  
 

\bibitem{Liddle:1994dx}
  A.~R.~Liddle, P.~Parsons and J.~D.~Barrow,
  Phys.\ Rev.\ D {\bf 50} (1994) 7222
  doi:10.1103/PhysRevD.50.7222
  [astro-ph/9408015].

\bibitem{Liddle:2000cg} A.~R.~Liddle and D.~H.~Lyth, 
\textit{Cosmological Inflation and Large-Scale Structure}, 
Cambridge University Press, (2000).

\bibitem{Ade:2015xua}
  P.~A.~R.~Ade {\it et al.} [Planck Collaboration],
  arXiv:1502.01589 [astro-ph.CO].

  
\bibitem{Lewis:1999bs}
  A.~Lewis, A.~Challinor and A.~Lasenby,
  Astrophys.\ J.\  {\bf 538} (2000) 473
  doi:10.1086/309179
  [astro-ph/9911177].
  
\bibitem{Lewis:2002ah}
  A.~Lewis and S.~Bridle,
  Phys.\ Rev.\ D {\bf 66} (2002) 103511
  doi:10.1103/PhysRevD.66.103511
  [astro-ph/0205436].
  
\bibitem{Ade:2015tva}
  P.~A.~R.~Ade {\it et al.} [BICEP2 and Planck Collaborations],
``Joint Analysis of BICEP2/Keck  Array and Planck Data,''
  Phys.\ Rev.\ Lett.\  {\bf 114} (2015) 101301
  doi:10.1103/PhysRevLett.114.101301
  [arXiv:1502.00612 [astro-ph.CO]].
  
\bibitem{German:2015qjq}
  G.~German, A.~Herrera-Aguilar, J.~C.~Hidalgo and R.~A.~Sussman,
  arXiv:1512.03105 [astro-ph.CO].

\bibitem{Kohri:2007qn}
  K.~Kohri, D.~H.~Lyth and A.~Melchiorri,
  JCAP {\bf 0804} (2008) 038
  doi:10.1088/1475-7516/2008/04/038
  [arXiv:0711.5006 [hep-ph]].
  
\bibitem{Josan:2009qn}
  A.~S.~Josan, A.~M.~Green and K.~A.~Malik,
  Phys.\ Rev.\ D {\bf 79} (2009) 103520
  doi:10.1103/PhysRevD.79.103520
  [arXiv:0903.3184 [astro-ph.CO]].


\bibitem{Carr:2009jm}
  B.~J.~Carr, K.~Kohri, Y.~Sendouda and J.~Yokoyama,
  Phys.\ Rev.\ D {\bf 81} (2010) 104019
  doi:10.1103/PhysRevD.81.104019
  [arXiv:0912.5297 [astro-ph.CO]].

\bibitem{Clesse:2010iz}
  S.~Clesse,
  Phys.\ Rev.\ D {\bf 83} (2011) 063518
  doi:10.1103/PhysRevD.83.063518
  [arXiv:1006.4522 [gr-qc]].
  
\bibitem{Clesse:2013jra}
  S.~Clesse, B.~Garbrecht and Y.~Zhu,
  Phys.\ Rev.\ D {\bf 89} (2014) 6,  063519
  doi:10.1103/PhysRevD.89.063519
  [arXiv:1304.7042 [astro-ph.CO]].
  
\bibitem{Kodama:2011vs}
  H.~Kodama, K.~Kohri and K.~Nakayama,
  Prog.\ Theor.\ Phys.\  {\bf 126} (2011) 331
  doi:10.1143/PTP.126.331
  [arXiv:1102.5612 [astro-ph.CO]].
  
\bibitem{Clesse:2012dw}
  S.~Clesse and B.~Garbrecht,
  Phys.\ Rev.\ D {\bf 86} (2012) 023525
  doi:10.1103/PhysRevD.86.023525
  [arXiv:1204.3540 [hep-ph]].
  
\bibitem{Mulryne:2011ni}
  D.~Mulryne, S.~Orani and A.~Rajantie,
  Phys.\ Rev.\ D {\bf 84} (2011) 123527
  doi:10.1103/PhysRevD.84.123527
  [arXiv:1107.4739 [hep-th]].
  
  

  
\bibitem{Carr:1974nx}
  B.~J.~Carr and S.~W.~Hawking,
  Mon.\ Not.\ Roy.\ Astron.\ Soc.\  {\bf 168} (1974) 399.
  
\bibitem{Khlopov:2008qy}
  M.~Y.~Khlopov,
  Res.\ Astron.\ Astrophys.\  {\bf 10} (2010) 495
  doi:10.1088/1674-4527/10/6/001
  [arXiv:0801.0116 [astro-ph]].
  
\bibitem{Frampton:2010sw}
  P.~H.~Frampton, M.~Kawasaki, F.~Takahashi and T.~T.~Yanagida,
  JCAP {\bf 1004} (2010) 023
  doi:10.1088/1475-7516/2010/04/023
  [arXiv:1001.2308 [hep-ph]].
  
\bibitem{Blais:2002nd}
  D.~Blais, C.~Kiefer and D.~Polarski,
  Phys.\ Lett.\ B {\bf 535} (2002) 11
  doi:10.1016/S0370-2693(02)01803-8
  [astro-ph/0203520].
  
\bibitem{Clesse:2015wea}
  S.~Clesse and J.~García-Bellido,
  Phys.\ Rev.\ D {\bf 92} (2015) 2,  023524
  doi:10.1103/PhysRevD.92.023524
  [arXiv:1501.07565 [astro-ph.CO]].
  
\bibitem{Kawasaki:2015ppx}
  M.~Kawasaki and Y.~Tada,
  arXiv:1512.03515 [astro-ph.CO].
\bibitem{Lyth:2010zq}
  D.~H.~Lyth,
  JCAP {\bf 1107} (2011) 035
  doi:10.1088/1475-7516/2011/07/035
  [arXiv:1012.4617 [astro-ph.CO]].
  
\bibitem{Lyth:2011kj}
  D.~H.~Lyth,
  arXiv:1107.1681 [astro-ph.CO].
  
\bibitem{Lyth:2012yp}
  D.~H.~Lyth,
  JCAP {\bf 1205} (2012) 022
  doi:10.1088/1475-7516/2012/05/022
  [arXiv:1201.4312 [astro-ph.CO]].
  
\bibitem{Bugaev:2011qt}
  E.~Bugaev and P.~Klimai,
  JCAP {\bf 1111} (2011) 028
  doi:10.1088/1475-7516/2011/11/028
  [arXiv:1107.3754 [astro-ph.CO]].
  
\bibitem{Bugaev:2011wy}
  E.~Bugaev and P.~Klimai,
  Phys.\ Rev.\ D {\bf 85} (2012) 103504
  doi:10.1103/PhysRevD.85.103504
  [arXiv:1112.5601 [astro-ph.CO]].


\end{thebibliography}
\end{document}